\documentclass{SciPost}

\hypersetup{
    colorlinks,
    linkcolor={red!50!black},
    citecolor={blue!50!black},
    urlcolor={blue!80!black}
}

\usepackage[bitstream-charter]{mathdesign}
\DeclareSymbolFont{usualmathcal}{OMS}{cmsy}{m}{n}
\DeclareSymbolFontAlphabet{\mathcal}{usualmathcal}

\fancypagestyle{SPstyle}{
\fancyhf{}
\lhead{\colorbox{scipostdeepblue}{\bf \color{white} ~SciPost Physics Proceedings }}
\rhead{{\bf \color{scipostdeepblue} ~Submission }}

\fancyfoot[C]{\textbf{\thepage}}
}

\begin{document}

\pagestyle{SPstyle}

\begin{center}{\Large \textbf{\color{scipostdeepblue}{
Interdisciplinary Digital Twin Engine InterTwin for calorimeter simulation \\
}}}\end{center}

\begin{center}\textbf{
Corentin Allaire\textsuperscript{1},
Vera Maiboroda\textsuperscript{1$\star$} and
David Rousseau\textsuperscript{1$\dagger$}
}\end{center}

\begin{center}
{\bf 1} CNRS, IJCLab
\\[\baselineskip]
$\star$ \href{mailto:email1}{\small vera.maiboroda@cern.ch}\,,\quad
$\dagger$ \href{mailto:email2}{\small david.rousseau@ijclab.in2p3.fr}
\end{center}

\definecolor{palegray}{gray}{0.95}
\begin{center}
\colorbox{palegray}{
  \begin{tabular}{rr}
  \begin{minipage}{0.37\textwidth}
    \includegraphics[width=60mm]{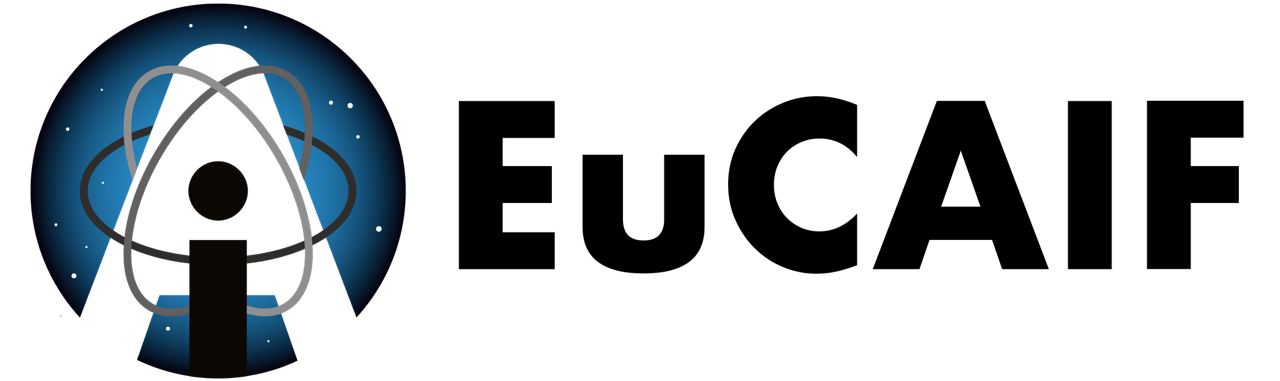}
  \end{minipage}
  &
  \begin{minipage}{0.5\textwidth}
    \vspace{5pt}
    \vspace{0.5\baselineskip} 
    \begin{center} \hspace{5pt}
    {\it The 2nd European AI for Fundamental \\Physics Conference (EuCAIFCon2025)} \\
    {\it Cagliari, Sardinia, 16-20 June 2025
    }
    \vspace{0.5\baselineskip} 
    \vspace{5pt}
    \end{center}
    
  \end{minipage}
\end{tabular}
}
\end{center}

\section*{\color{scipostdeepblue}{Abstract}}
\textbf{\boldmath{%
Calorimeter shower simulations are computationally expensive, and generative models offer an efficient alternative. However, achieving a balance between accuracy and speed remains a challenge, with distribution tail modeling being a key limitation. Invertible generative network CaloINN provides a trade-off between simulation quality and efficiency. The ongoing study targets introducing a set of post-processing modifications of analysis-level observables aimed at improving the accuracy of distribution tails. As part of interTwin project initiative developing an open-source Digital Twin Engine, we implemented the CaloINN within the interTwin AI framework.
}}

\vspace{\baselineskip}

\noindent\textcolor{white!90!black}{%
\fbox{\parbox{0.975\linewidth}{%
\textcolor{white!40!black}{\begin{tabular}{lr}%
  \begin{minipage}{0.6\textwidth}%
    {\small Copyright attribution to authors. \newline
    This work is a submission to SciPost Phys. Proc. \newline
    License information to appear upon publication. \newline
    Publication information to appear upon publication.}
  \end{minipage} & \begin{minipage}{0.4\textwidth}
    {\small Received Date \newline Accepted Date \newline Published Date}%
  \end{minipage}
\end{tabular}}
}}
}



\section{Introduction}
\label{sec:intro}

Modern particle and nuclear physics experiments rely heavily on detailed simulations to connect theoretical predictions with experimental measurements. As the Large Hadron Collider (LHC) enters new data-taking phases, the demand for simulated samples will rise, placing significant pressure on available computing resources. Within the ATLAS experiment \cite{atlas} at LHC, a substantial share of this demand comes from the simulation of the detector responses, with calorimeter simulations being especially resource-intensive. The challenge lies in accurately describing the complex particle showers consisting of many secondary particles produced when particles interact with dense calorimeter materials. Geant4 \cite{geant4} remains the standard for realistic detector simulations, but its heavy cost in computing time raises serious challenges for future experiments.

As the LHC continues to deliver ever-increasing amounts of data, the ability to generate matching simulated samples has become a central challenge for precision measurements. If the production of Monte Carlo events cannot keep pace, analyses risk being limited not by the data itself but by the statistical reach of the simulations. To address this, faster approaches to calorimeter modeling have been developed, often using parametrized detector responses to replace the detailed shower calculations of Geant4. While such strategies drastically reduce computational requirements, they generally fall short in accuracy for the most demanding measurements. Recent advances in machine learning, particularly in generative models, offer a promising alternative. By learning the underlying distributions of particle interactions, deep generative models can provide fast, high-quality event simulations, and have already begun to be integrated into experimental toolkits as potential successors to traditional fast-simulation methods.

The relevance of simulation and digital-twin technologies extends beyond high-energy physics. Digital twins are increasingly applied in fields such as climate science, engineering, and materials research, where they enable predictive modeling and the integration of heterogeneous data. The interTwin project \cite{intertwin} addresses this broader context by developing a Digital Twin Engine (DTE), a co-designed, open-source framework based on modular components and open standards, to support interoperable, cross-domain digital-twin applications. Among the targeted applications is high-energy physics, where interTwin supports the development of data-driven calorimeter detector twins to accelerate simulation and analysis workflows.

In this work, we adopt a deep generative model CaloINN \cite{caloinn} for fast calorimeter simulation (Section \ref{sec:caloinn}) and implement it within the interTwin AI framework (Section \ref{sec:intertwin}). We further investigate strategies to enhance the performance and fidelity of this generative approach, discussed in Section \ref{sec:tails}.

\section{Choosing a model for fast calorimeter simulation}
\label{sec:caloinn}

The Fast Calorimeter Simulation Challenge (CaloChallenge) \cite{calochallenge} was launched in 2022 had a goal to stimulate progress in fast calorimeter simulation and to establish a common ground for benchmarking. The initiative invited participants to train generative models on dedicated calorimeter-shower datasets, with the task of learning the conditional distribution 
$p(I|E_{inc})$, where $I$ denotes voxelized energy deposits and $E_{inc}$ the incident particle energy. By providing standardized datasets and a unified evaluation pipeline, the challenge sought to advance existing fast calorimeter shower simulation efforts while enabling consistent comparison across approaches.

Three datasets of increasing complexity were released. The first, derived from ATLAS open data, contains Geant4-based simulations of single photons and charged pions injected at the calorimeter surface and pointing to the detector center. The resulting showers offer idealized inputs for algorithm training, with 368 and 533 voxels per sample for photons and pions, respectively. The remaining two datasets were produced with the Geant4 Par04 example, modeling an idealized cylindrical calorimeter with alternating absorber and active layers, and feature higher granularity, with 6480 and 40,500 voxels correspondingly.

More than 30 generative models were submitted to CaloChallenge, covering broad spectrum of existing generative model architectures. As expected, no single approach dominated: methods emphasizing fidelity, such as diffusion and transformer-based models, achieved higher quality samples but required longer generation times, whereas GAN- and VAE-based approaches were faster but less accurate. Notably, normalizing-flow models achieved strong performance across both speed and simulation quality. 

Motivated by the strong performance of flow-based models in the CaloChallenge, we chose to base our studies on CaloINN. CaloINN is an invertible neural network employing coupling layer-based normalizing flows. This type of networks are equally fast to evaluate in both directions, making training and generation more efficient than in autoregressive setups. The bijective transformation in CaloINN is based on spline functions, allowing for expressive and flexible modeling of calorimeter showers. To enhance performance on a high-granular calorimeter geometries, a Variational Autoencoder (VAE) is used to compress the dimensionality of the data before applying the normalizing flow.

\section{interTwin project}
\label{sec:intertwin}

The interTwin project is a European Union–funded initiative that aims to develop a cross-disciplinary Digital Twin Engine (DTE), providing an open-source modular platform for the creation and deployment of scientific digital twins. Its architecture combines core infrastructure modules with domain-specific thematic modules. A central component of this ecosystem is itwinai, a Python-based framework designed to streamline the use of artificial intelligence in digital-twin applications. It offers experiment tracking and model registry through MLFlow \cite{Zaharia_Accelerating_the_Machine_2018}, support for distributed training across HPC and cloud environments, and workflow definition via configuration files. These features reduce the engineering burden associated with machine-learning workflows.

Among the many scientific use cases, the high-energy physics application focuses on accelerating calorimeter simulations through generative models. Within this use case, two models are available. The first is 3DGAN \cite{3dgan}, a generative adversarial network that produces three-dimensional calorimeter showers represented as voxelized energy depositions in calorimeters. Its generator is conditioned on the incident particle energy and the angle of incidence. By exploiting three-dimensional convolutional architectures, 3DGAN achieves speed-ups of several orders of magnitude over Geant4 while preserving key calorimetric observables such as total energy deposition and longitudinal and lateral shower profiles. The second model, CaloINN, has also been integrated into the interTwin framework. Within itwinai, the implementation benefits from configuration management that consolidates all data, model, and training parameters into a single configuration file. This includes access and setup of experiment tracking via MLFlow and distributed computing strategies. Such approach ensures reproducibility and facilitates large-scale experiments. The presence of both 3DGAN and CaloINN in the interTwin ecosystem demonstrates the feasibility of applying the platform to high-energy physics simulations.

\section{Fixing distribution tails}
\label{sec:tails}

\begin{figure}
    \centering
    \includegraphics[width=0.6\linewidth]{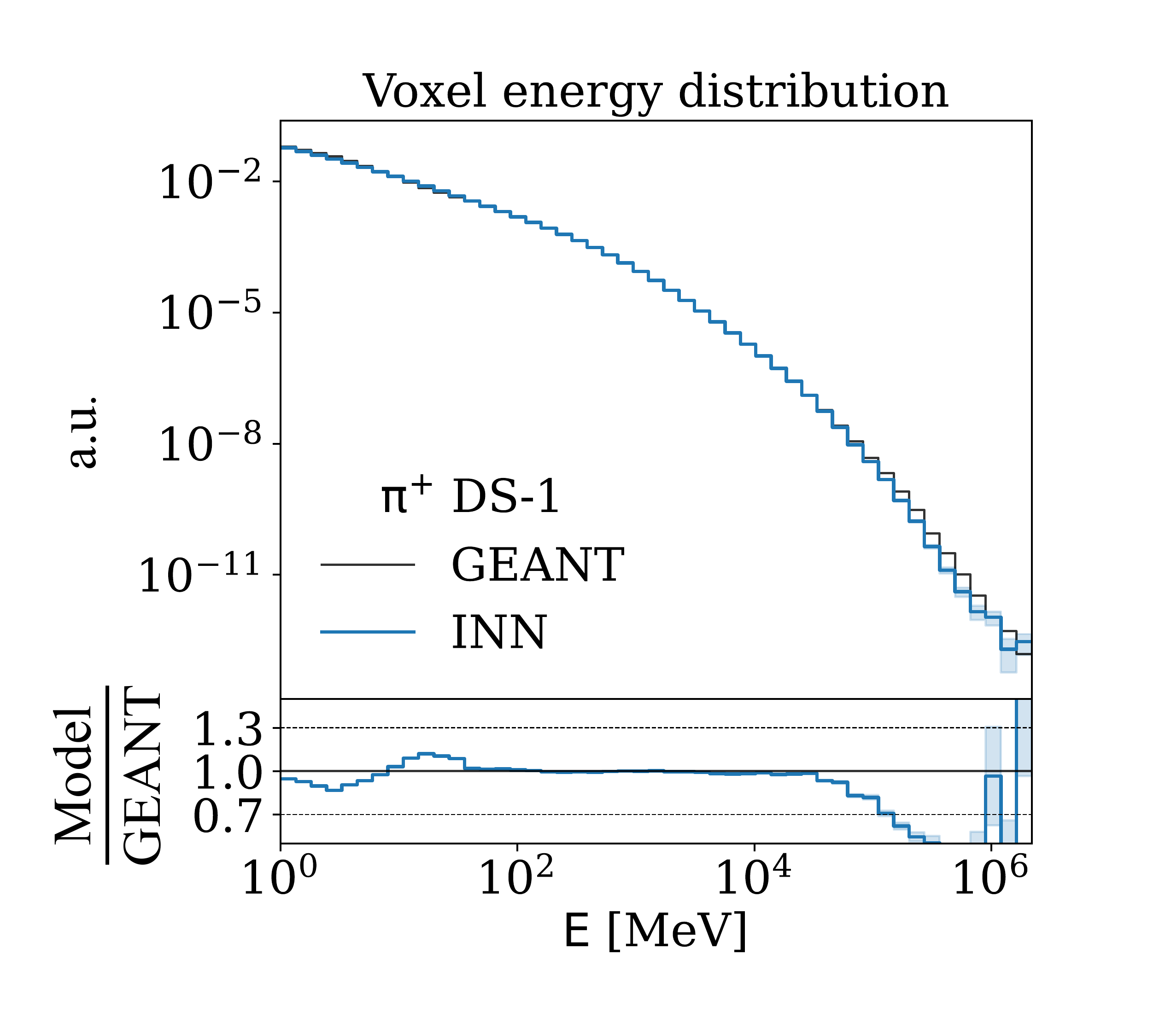}
    \caption{Distribution of voxel energy compared between Geant4 (black) and CaloINN (blue) on the CaloChallenge dataset containing pions. While the bulks of distributions are in a good agreement, some discrepancy can be observed in the tails.}
    \label{fig:caloinn_voxelen}
\end{figure}

The central objective of the ongoing research is to improve the fidelity of the CaloINN model. In practice, this requires achieving accuracies better than 10\% in the distribution tails at the $10^{-3}$ quantile level. Achieving this level of precision is essential for ensuring that fast simulation tools do not bias rare-event searches or precision measurements. Without any corrections, the performance of the CaloINN model is shown on Figure \ref{fig:caloinn_voxelen}.

Recent work, such as the Fast Perfekt approach \cite{fastperfekt}, has demonstrated the potential of refining fast simulations through lightweight machine-learning corrections. Fast Perfekt employs a regression-based residual neural network that learns deterministic morphing functions to align fast simulation outputs with full simulation targets. This illustrates that targeted refinement strategies can bring fast simulations closer to Geant4-like accuracy without sacrificing their efficiency.

Building on this idea, our ongoing work explores fine-tuning procedures at the inference stage. CaloINN is first to be trained on datasets with artificially enhanced tails, thereby over-representing extreme calorimeter showers. Such showers can be selected based on anomaly scores computed using Isolation Forest \cite{isolforest} or Local Outlier Factor \cite{lof} algorithms, which quantify how strongly a given shower deviates from the rest of the dataset. A dedicated classifier is then to be introduced to distinguish between generative model outputs and Geant4-based simulations, enabling the estimation of density ratios across the feature space. These ratios are subsequently applied during inference to reweight and correct the generated distributions, pushing the tails toward the Geant4 reference. The concrete implementation of this procedure is currently under investigation. This approach draws inspiration from variance-reduction techniques \cite{Bruenner_2021} and is designed to preserve the overall speed advantage of generative models. The additional computational cost is expected to remain below a factor of two, which is acceptable given that the baseline speed-up relative to Geant4 typically spans two to three orders of magnitude.

\section{Conclusion}
Accurate and efficient detector simulation remains a central challenge for high-energy physics in the high-luminosity era of the LHC. In this work, we are situating the developments of improved calorimeter simulation techniques within the broader context of the interTwin project. The integration of both 3DGAN and CaloINN into interTwin demonstrates the viability of the platform for high-energy physics applications. We are pursuing strategies to bring generative models closer to production readiness by improving tail fidelity without compromising their computational advantage.

\section*{Acknowledgements}

\paragraph{Funding information}
interTwin is funded by the European Union Grant Agreement Number 101058386.

\begin{appendix}

\end{appendix}







\bibliography{SciPost_Example_BiBTeX_File.bib}


\end{document}